\documentstyle[prb,aps,epsfig,twocolumn]{revtex}
\def\bmf#1{\boldmath$\bf\displaystyle#1$\unboldmath}

 \draft
\title { \bf Evidence of anisotropic magnetic polarons in La$_{0.94}$Sr$_{0.06}$MnO$_3$  by neutron scattering and comparison with Ca-doped manganites}
\author{M. Hennion$^1$, F. Moussa$^1$, G. Biotteau$^1$, J. Rodr\'{\i}guez-Carvajal$^1$,
L. Pinsard$^2$ and A. Revcolevschi$^2$ }

\address{1 - Laboratoire L\'eon Brillouin, CEA-CNRS, CE Saclay, 
91191 Gif sur Yvette Cedex France}
\address{2 - Laboratoire de Physico-Chimie des Solides 
Universit\'e Paris-Sud 91405 Orsay Cedex France}
\date{ Phys. Rev. B Volume 61, Number 13, 1 April 2000}
\begin{document}
\twocolumn[\hsize\textwidth\columnwidth\hsize\csname @twocolumnfalse\endcsname
\maketitle
\begin{abstract}
Elastic and inelastic neutron scattering experiments have
 been performed in a La$_{0.94}$Sr$_{0.06}$MnO$_3$ untwinned crystal, which exhibits 
an antiferromagnetic canted magnetic structure with ferromagnetic layers.
 The elastic small q scattering exhibits a modulation with an anisotropic q-dependence. It can be pictured by ferromagnetic inhomogeneities or large magnetic polarons with a platelike shape, the largest size ($\approx$17$\AA$)
 and largest inter-polaron distance ($\approx$ 38$\AA$) being within the ferromagnetic layers. Comparison with observations performed on Ca-doped samples, which show the growth of the magnetic polarons with doping, suggests that this growth is faster for the Sr than for the Ca substitution. Below the gap of the spin wave branch typical of the AF layered magnetic structure, 
an additional spin wave branch reveals a ferromagnetic and isotropic coupling, already found in Ca-doped samples.
Its q-dependent intensity, very anisotropic,  
closely reflects the ferromagnetic correlations found for the static clusters.  All these results agree
with a two-phase electronic segregation occurring on a very small scale, although some characteristics of a canted state are also observed suggesting a weakly inhomogeneous state. 
\end{abstract}

]
\section{INTRODUCTION}

In the simple ionic picture of La$_{1-x}$B$_x$MnO$_3$ (B=Sr$^{2+}$, Ca$^{2+}$, or Ba$^{2+}$),  upon increasing x, the Mn$^{3+}$ ions characterized by three t$_{2g}$ and one e$_g$ electrons are 
gradually replaced by Mn$^{4+}$  with empty 
e$_g$ orbitals. These compounds have attracted much attention due to their anomalous transport, 
magnetic and structural properties. The low-doping range of these compounds preceding the ferromagnetic and 
metallic phase is far from being fully 
explored, and the mechanism driving the metal-insulator transition is still unclear. On 
the theoretical side, the situation is rather controversial. Two pictures have been proposed. 
In the model of de Gennes\cite{de Gennes}, the competition between the AF superexchange and the ferromagnetic double-exchange couplings associated
with the hole hopping\cite{Zener}, results in a canted state. In contrast with this
 picture, several authors have found 
 that an inhomogeneous representation of  ferromagnetic hole-rich regions (called ferrons or ferromagnetic polarons), within an 
antiferromagnetic hole-poor medium, could be more favorable than that of a canted state\cite{Nagaev,Gor'kov,Kagan,Arovas}. 
Similar results have been obtained by Monte-Carlo simulations\cite{Yunoki,Moreo,Yi}.  This picture of ferromagnetic polaron, here, differs from the spin polaron picture discussed  by Kasuya\cite{Kasuya}, which consists of a ferromagnetic cloud surrounding one hole.

In La$_{1-x}$Ca$_x$MnO$_3$, interesting observations have been recently 
reported: a "static" ferromagnetic modulation has been observed at small wave-vectors q, 
interpreted in terms of magnetic inhomogeneities or droplets, with
a distribution analogous to that of a liquid\cite{Hennion2}. 
Moreover, below the gap of the spin wave branch with an anisotropic dispersion typical of the AF layered structure, an
additional spin wave branch with an isotropic dispersion was 
observed\cite{Hennion2,Hennion1,Moussa2}. Whether these findings
  are  restricted to Ca
  substitution, in relation with the stronger induced lattice distortion, or are more general, is 
an important question. The present study carried out on Sr-substituted samples sheds light on this issue. Sr is known to induce a smaller distortion of the  
Mn-O-Mn bonds than Ca, resulting in a higher hopping probability, t, for the e$_g$ holes. Therefore, 
in the ferromagnetic and metallic state, at large x values, the ferromagnetic transition temperature 
is much higher for Sr than for Ca substitution.
This enhanced ferromagnetic character can also be
 seen at smaller x, by a shift of the critical concentration x$_c$ for the metal-insulator transition. 
For instance, the x$_{Sr}$=0.175 sample is metallic\cite{Mukhin,Moritomo}, whereas the x$_{Ca}$=0.194 one is still insulating\cite{Joonghoe Dho}. Previous inelastic
 neutron measurements on low doped La$_{1-x}$Sr$_x$MnO$_3$, with x=0.05\cite{Hirota} and 
0.09\cite{Moudden}, have reported a single spin wave branch, with AF interplane and F in-plane magnetic couplings.

The present study, carried out on an untwinned x$_{Sr}$=0.06 sample, reveals two additional 
features. Below the temperature of the magnetic transition, a ferromagnetic modulation
of the intensity of neutron scattering can be observed at small q, similarly to our previous observations in one Ca-substituted sample\cite{Hennion2}. The untwinned sample reveals the anisotropic character of this modulation. 
The size of the inhomogeneities is about twice as larger (17$\AA$) within the ferromagnetic layers as perpendicularly
 to them, indicating platelets instead of droplets. Furthermore, an additional spin
 wave branch is observed with an isotropic dispersion curve, 
characteristic of a ferromagnetic coupling. The q dependence of the corresponding susceptibility, along the two main symmetry directions, reflects closely the correlations of the static inhomogeneities, showing the same anisotropy. This study confirms the general character 
of our previous observations. The density of the platelets which is much lower than the hole density
confirms a picture of an electronic phase segregation on the scale of two tens of $\AA$.  
Comparison with the previously studied Ca-substituted sample (x$_{Ca}$=0.08) and with a new one with a smaller x$_{Ca}$ value (x$_{Ca}$=0.05) reported here , brings up to deduce that the ferromagnetism induced by hole doping is enhanced upon Sr substitution. Such an enhancement is observed  both in the size of the ferromagnetic polarons and in the characteristics of the spin 
dynamics. 

The rest of the paper is organized as follows: In section II the magnetic structure is determined. In section III A and B, elastic diffuse scattering data at small angle and 
around the
(110) Bragg peak are reported. In section IV, the spin dynamics is investigated. Finally, in section V, the 
overall results are discussed. 

\section{Magnetic structure}

    Experiments have been 
carried out on the three-axis spectrometer 4F1, installed on a cold source at the reactor
 Orph\'ee (Laboratoire L\'eon Brillouin), on an untwinned single crystal of La$_{0.94}$Sr$_{0.06}$MnO$_3$ 
(0.1 cm$^3$), grown by a floatting zone method.
The nominal 
x=0.06 value agrees with x deduced from the lattice parameters reported in\cite{Urushibara}\cite{Kawano}. The structure is 
orthorhombic
 ({\bf $c/\sqrt 2<a<b$})
with $Pbnm$ symmetry. 
The temperature dependence of (110),
(001) and (112) Bragg-peak intensities was studied. 
The $\tau$=(001) Bragg-peak intensity determines a long range antiferromagnetic order below T$_N$=130K (see Fig. 1-a). 
The  evolution of the $\tau$=(112) Bragg peak intensity, 
very sensitive to the occurence of ferromagnetism because of its small nuclear intensity, determines 
a long range ferromagnetic order at T$_K$=124 K, smaller than T$_N$ (see Fig 1-b, c). 
The absence of any magnetic component below T$_K$
 in the (002) Bragg peak intensity, indicates that the ferromagnetic spin component is along [001] 
(geometrical factor). 
Therefore 
an {\it average} canted state is defined below T$_K$, which consists of ferromagnetic layers in the ({\bf a, b}) plane, antiferromagnetically stacked along {\bf c}, and with a weak ferromagnetic 
spin component along {\bf c} as sketched in Fig. 1-d.
 The two transition temperatures agree with the phase diagram of Ca-doped 
samples\cite{Moussa2}
being between those 
found for x$_{Ca}$=0.05 and  x$_{Ca}$=0.08.
 However, an interesting difference between the two systems appears for the spin deviation 
or canting angle $\theta$ deduced from the weak ferromagnetic component. From the (112) Bragg peak intensity, an average deviation angle 
$\theta$
 from {\bf b} towards {\bf c} of 13$^0$ is determined at T=15K; this value is the same as that found for x$_{Ca}$=0.08,  and much larger than the value, $\theta$=5$^0$,
 determined for x$_{Ca}$=0.05. Therefore, the increase of the canted angle $\theta$, upon increasing x, is faster for a Sr substitution than for a Ca one.
It is worth-noting that the Bragg-peak intensities define an average magnetization ${\bar m}$ but 
do not inform 
about its
 homogeneous or inhomogeneous character. The inhomogeneities contribute to diffuse scattering 
between Bragg peaks which is studied below.

\begin{figure}
\centerline{\epsfig{file=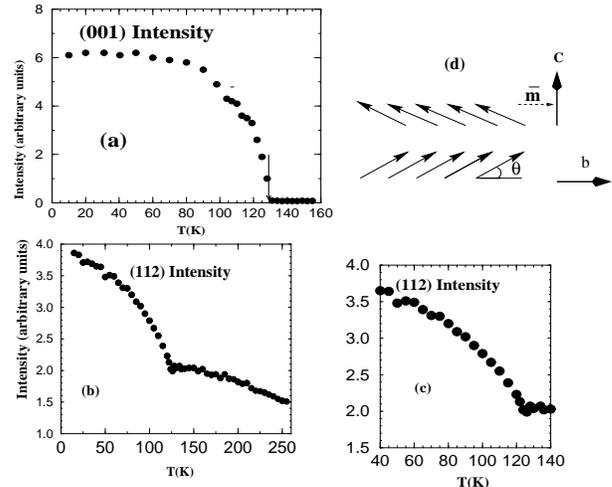,width=8cm}}
\caption{{\bf a)} Temperature variation of the intensity of {\bf (a)} the (001) Bragg peak defining 
T$_N$ (130K) , and {\bf (b, c)} the (112) Bragg peak defining T$_K$= (124K). In (d), the
 drawing represents the average magnetic structure.
 }
\end{figure}

\section{Elastic diffuse scattering}

Elastic ($\hbar\omega$=0) diffuse scattering measurements, which give access to short-range ordered correlations, have been performed at several 
temperatures along several {\bf Q}-paths in the reciprocal lattice, with {\bf Q}={\bf q}+\bmf{\tau}.
In section A, diffuse scattering intensity close to the direct 
beam \bmf{\tau}=(000) is reported along
the main symmetry directions.
In section B, diffuse scattering around \bmf{\tau}=(110) is reported along the same q 
directions.  
 
These two types of experiments appear complementary. The scattering intensity close to the direct 
beam is more easily modelized, but, because of the geometrical factor, the [001] symmetry 
direction is unobservable. In contrast, this direction
can be observed close to \bmf{\tau}=(110). We also compare the diffuse scattering close to \bmf{\tau}=(110)
 with that measured close to \bmf{\tau}=(220), where, due to the Mn magnetic form factor, 
the magnetic intensity is expected to decrease. 
Finally, antiferromagnetic correlations have also been measured close to \bmf{\tau}=(001)
along the [001] direction and reported in section C.

\subsection{Ferromagnetic short-range order studied close to \bmf{\tau}=(000)}
1) {\it Experiment}

 \begin{figure}
\centerline{\epsfig{file=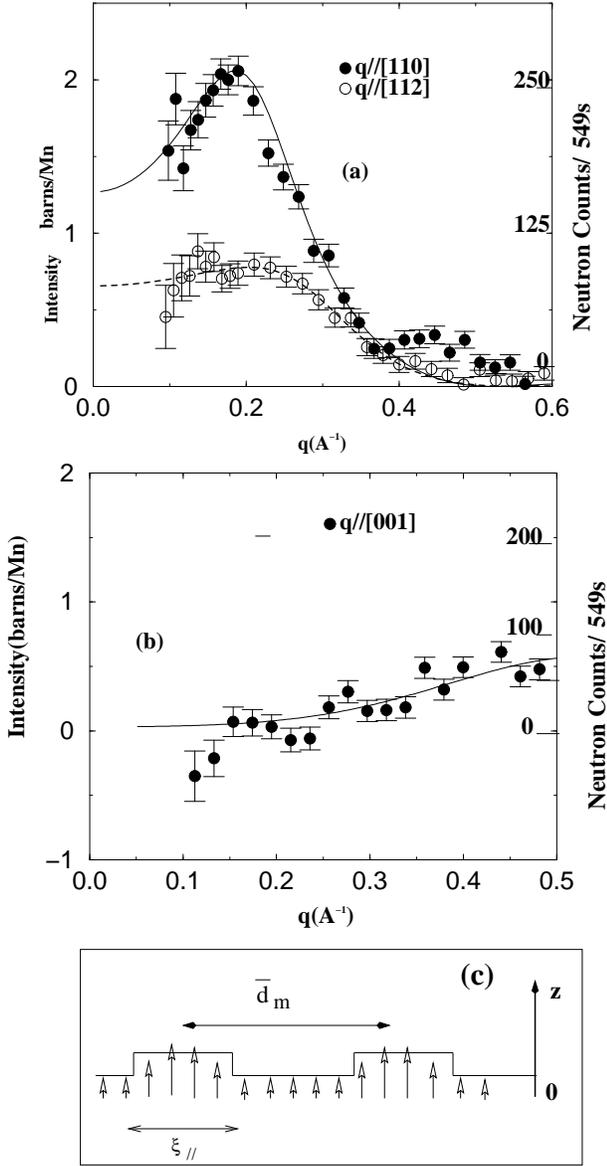,width=8cm}}
\caption{ Elastic magnetic scattering I$_{15K}$-I$_{HT}$ versus q in barns per Mn (left hand scale)
 and neutron counts (right hand scale): {\bf (a)} along [110] and [112] q directions and {\bf (b)} 
along [001] q direction. The lines are guides to the eye. {\bf (c)}: schematic representation  of a modulation of longitudinal components S$^z$.}
\end{figure}

Small angle scattering experiments have been performed in the range 0.1$\AA^{-1}$$<$q$<$0.6$\AA^{-1}$ 
, and in the temperature range 15K$<$T$<$300K along [110], [112] and [001] directions using k$_i$=1.25$\AA^{-1}$.
 
As in\cite{Hennion2}, the magnetic intensity I(q) was obtained by 
subtracting the high temperature spectrum, I$_{HT}$(q), found to be q-isotropic and temperature independent for T$>$T$_{K}$, from the spectrum measured at 15K. Such a subtraction procedure assumes that  I$_{HT}$(q) has a 
nuclear 
origin assigned to dislocations. Actually a residual scattering intensity above T$_K$ can be observed close to \bmf{\tau}=(110) where dislocations do not 
contribute (sub-section B). Since it is very small, the subtraction procedure is valid. I(q) was converted in barns per Mn 
using a Vanadium standard, 
and is reported in Fig 2-a for q along [110] and [112] and in Fig 2-b along [001]. In both [110] 
and [112] directions, a modulation is observed
 around q=0.2$\AA^{-1}$, differing in intensity. The weak maximum observed at larger q along [110], since not observed close to the equivalent symmetry point (110), is likely a spurious effect.
 The magnetic contribution observed along [001], is nearly zero around  0.2$\AA^{-1}$, and 
increases slightly with q (Fig 1-b).

2) {\it Analysis}

In our previous study of a Ca-doped sample\cite{Hennion2}, the apparent isotropy was the consequence of twinned domains. We have proposed a crude model of magnetic inhomogeneities 
or droplets. 
The basic assumption is that the 
scattering intensity results from the contribution of longitudinal spin components S$^z$ along [001] as sketched in Fig. 2-c. In addition, we have assumed that the characteristics of the magnetic inhomogeneities  (shape, distance) are the same whatever the 
direction (isotropic model). 
 Our untwinned sample provides the opportunity to check  these two 
assumptions.
 By analogy with the chemical problem where the nuclear scattering length b$_{i,j}$ replaces S$^z_{i,j}$, the Fourier transform ${\cal I}$({\bf Q)} of the correlations of S$^z$ spin components may be written:

${\cal I}$({\bf Q})$\propto$$\sum_{i,j=1}^N$$ ({\bar S}^{z\perp})^2$$e^{i{\bf Q}({\bf R}_i-{\bf R}_j)}$+

$\sum_{i,j=1}^N$$(S^{z\perp}_iS^{z\perp}_j-({\bar S}^{z\perp})^2) e^{i{\bf Q}({\bf R}_i-{\bf R}_j)}$

 Here, the symbol $\perp$ means the projection of S$^z$ in the plane perpendicular to {\bf Q}. The first term gives rise to the ferromagnetic Bragg peaks and the second one to ferromagnetic short range order. The
bar stands for spatial average. The thermal average has been omitted. In the present case, {\bf Q}={\bf q}, since \bmf{\tau}=(000). By approximating the inhomogeneous character of the 
magnetization to a step function characterized by two magnetization values (see Fig 2-c),
and replacing the discrete spin S$_i$ by the continuous magnetization density m$^{\perp}_z$({\bf r}), the second term may 
be factorized into two functions $F(q,R)$ and $J(q, d_m, d_{min})$:

\begin{eqnarray}
 I(q)=C{\cal F}^2(Q)|F(q,R)|^2J(q, d_m, d_{min})                      
\end{eqnarray}

where $F(q,R)$ is the Fourier transform of one cluster assimilated to a sphere of radius 
$R$, and $J(q)$ 
 describes the cluster spatial distribution\cite{Guinier}. Its full expression which uses the average inter-cluster distance $d_m$ and a minimal distance of approach d$_{min}$, has been previously 
given\cite{Hennion2}. ${\cal F}$(Q) is the Mn magnetic form factor. Close to $\tau$=(000), ${\cal F}(Q)\approx 1$. From equation (1), mainly two length scales are derived: (i) the 
typical size
 of the inhomogeneity $R$ is mainly deduced from the q dependence beyond the maximum 
of I(q), since $J(q)$ in (1) tends asymptotically to unity,
  (ii) the characteristic distance between the inhomogeneities, $d_m$, (or equivalently, the 
inverse
 of the density $N_V ^{-1}$), deduced from the liquidlike distribution function $J(q, d_m, d_{min})$ 
in (1). An approximate determination can also be made from q$_{max}$ (q at maximum of
 intensity) with  $d_m$=2$\pi$/q$_{max}$. These two parameters  are included in the scale factor $C$ of (1), $C=(\gamma r_0)^2 |\Delta m^{z\perp}|^2 N_V$V$^2$, ($\gamma, r_0$ being constants for neutron scattering). The only free parameter when fitting is the squared magnetic 
contrast $|\Delta m^{z\perp}|^2$, given by the difference of the magnetizations between the two regions
(sketched in Fig. 2-c). In this model, the intensity is very sensitive to the size ($R$ enters at the power 6 in the constant $C$ for a spherical shape through the squared volume $V^2$). Therefore, in an isotropic model, the scattering 
intensity I(q) changes only by the geometrical factor as 
the q direction varies, decreasing by $\approx $1/2 from [110] (perpendicular to {\bf c}) to [112], down to zero along [001]. 

Comparison of these predictions with observations of Fig. 2-a and Fig. 2-b, leads to the following conclusions. The striking result is the absence of any intensity in the range 0.1$<$q$<$0.2 $\AA^{-1}$ along [001], where the maximum of intensity was observed in twinned samples. 
This confirms the assignment of this magnetic scattering to $S^z$ spin components. The increasing intensity at larger q indicates short-range antiferromagnetic correlations (the AF 
Bragg peak (001) is located at  q=0.83$\AA^{-1}$ in Fig 2). They will be discussed in section C.
Along [110] and [112], the scattering intensities do not only differ by a simple geometrical factor, 
but 
also by some anisotropy in their q dependence. The slower q dependence of I(q) beyond the 
maximum, for q along [112] indicates a smaller size along the [112] direction 
(at $\approx$ 45$^0$ from {\bf c}) suggesting a still smaller size along the [001]direction. 
Diffuse scattering experiments close to $\tau$=(110) reported in section B will characterize 
this anisotropy.

In the absence of any anisotropic model, a semi-quantitative description can be made, using the equation 
(1) to fit I(q) along the [110] direction only. The fit (continuous line in Fig 2-a) provides a 
correlation length $2R$=$\xi_{//}$$\approx$17$\AA$ and a density of clusters N$_V$ which corresponds to an inter-cluster distance which is twice the diameter value. The low density of the ferromagnetic clusters cannot be explained by the picture proposed for diluted chalcogenide alloys\cite{Kasuya} where a single electron ferromagnetically polarizes the neighbor spins
 but rather indicates an electronic phase separation.  In the rest of the paper, these ferromagnetic inhomogeneities are called ferromagnetic 
clusters
 or large ferromagnetic polarons in an equivalent way.

In Fig. 3, I(q) for x$_{Sr}$=0.06  is compared
 with  I(q) for x$_{Ca}$=0.08\cite{Hennion2} and x$_{Ca}$=0.05, in absolute units, for {\bf q} along the  [110] direction 
and at T=15K. For the
 Ca-substituted
 samples, a factor 2/3 takes into account the fact that one domain out of three does not 
contribute  (e.g. we report 3/2 I(q), I(q) being displayed in\cite{Hennion2}). We estimate a possible  
error of 10\%
 in the intensities, due to the fact that they are corrected for sample transmission which is calculated instead of 
measured.

 Considering first the two Ca-substituted samples, the analysis (dashed and dot-dashed lines)
shows the growth of the clusters with x$_{Ca}$ from $\approx 14\AA$ to $\approx 17\AA$ as x$_{Ca}$
 varies from 0.05 to 0.08. 
In addition, there is a small shift of q$_{max}$ with x$_{Ca}$, interpreted as an increase of the cluster density. The increase of the intensity by a factor two nearly agrees with equation (1) , being 30\%
smaller than expected. Interestingly, when comparing the x$_{Sr}$=0.06 sample to the two Ca-substituted ones, the maximum intensity of I(q) ($\approx$ 2 barns/Mn) and the q dependence 
beyond q$_{max}$ nearly coincide with the values of the x$_{Ca}$=0.08 case. Within the above model 
(continuous and dot-dashed lines), it indicates that the cluster size is the {\it same} in 
the two samples. We note a small shift of the q$_{max}$ value between both samples, also observed in the raw data around $\tau$=(110), reported in the inset of Fig. 3 in reduced lattice units ($rlu$),  which could indicate a slightly smaller cluster density in the x$_{Sr}$=0.06 sample. 

It results from the analysis of these three samples, that the intensity seems to be mainly determined by the size, in agreement with the model given by equation (1). 
This also indicates that the magnetic contrast value is the same in the three samples. 
As previously reported\cite{Hennion2}, the value deduced
 from the scale factor $C$ in (1), $ |\Delta m^{z\perp}|$$\approx $0.7$\mu_B$ is surprisingly 
small. This could result from an overestimate of the volume due to the shape anisotropy, the [110] direction corresponding to the largest size. Moreover, a slow decrease of the magnetization starting from the core of the polaron, as sketched in Fig. 2-c, is more realistic 
than the step function of the present model. We indicate that NMR experiments under an applied field suggest that some spins are fully ferromagnetic\cite{Allodi}.

In conclusion of the above analysis, 
 the size of the cluster, at a given x, depends on the chemical dopant, being larger for Sr than for Ca. This outlines the r\^ole of the  hopping
 probability "t", regulated by the mean Mn-O-Mn bond angle, to define the size.

\begin{figure}
\centerline{\epsfig{file=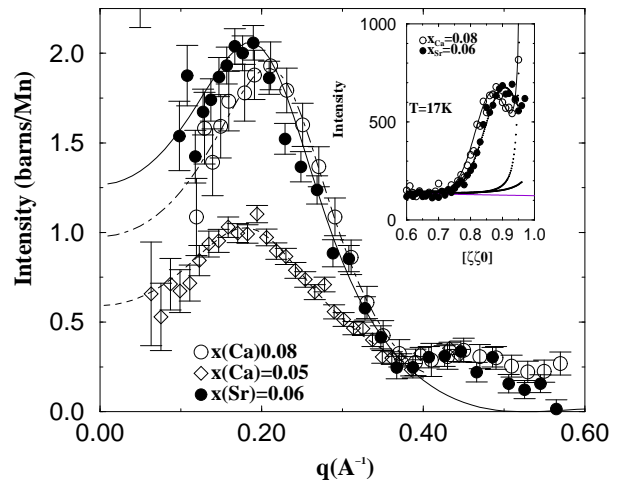,width=8cm}}
\caption{Magnetic scattering intensity versus q in barns/Mn. Full circles: x$_{Sr}$=0.06, open circles: x$_{Ca}$=0.08 and open squares: x$_{Ca}$=0.05. The continuous, dot-dashed and dashed
 lines are fits with the model (see the text). Inset: diffuse scattering (raw data) close to the (110) Bragg peak in equivalent reduced lattice units ($rlu$). Full circles: x$_{Sr}$=0.06, open circles: x$_{Ca}$=0.08. The continuous lines are guides for the eyes.
 }
\end{figure}

\subsection{Ferromagnetic short-range order studied close to  \bmf{\tau}=(110)}

Diffuse scattering intensity has been
 measured close to $\tau$=(110)
 for main symmetry directions and as a function of temperature, using the incident neutron 
wavevector k$_i$=1.55$\AA^{-1}$.

\begin{figure}
\centerline{\epsfig{file=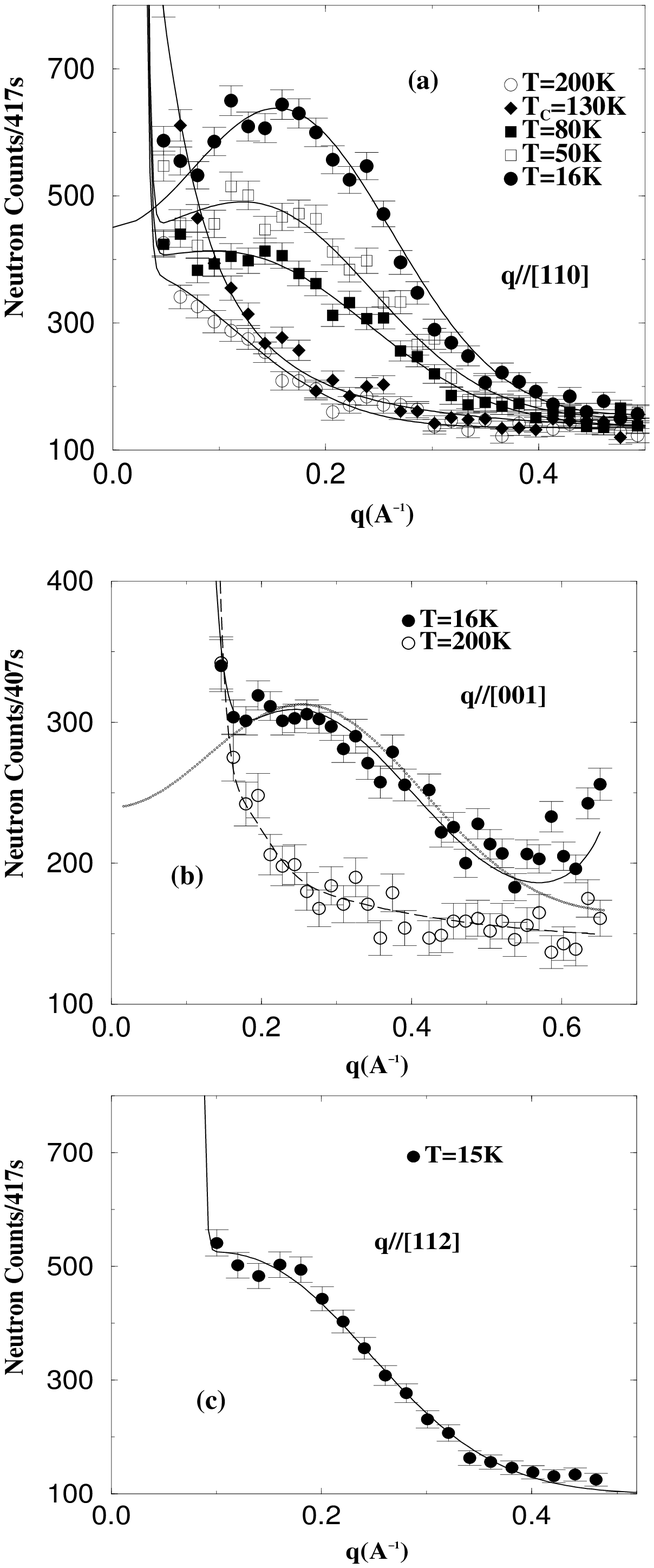,width=8cm}}
\caption{{\bf a}: Raw intensity versus {\bf q}, converted in $\AA^{-1}$. {\bf q} in {\it rlu} has been deduced from {\bf q}={\bf Q}-\bmf{\tau}, \bmf{\tau}=(110). {\bf a}: along [110], (0.4,0.4,0)$<${\bf Q}$<$(1,1,0). {\bf b}: along [001], (1,1,0)$<${\bf Q}$<$(1,1,0.85) and  {\bf c}: along [112], (1,1,0)$<${\bf Q}$<$ (1.3,1.3,0.6). The continuous lines are fits with a gaussian (modulation ) and delta functions (Bragg peak) except at 130K in {\bf (a)}
 where a lorentzian function is used. In {\bf b}, the gaussian component, corrected for the geometrical factor, is shown by a dotted curve.}
\end{figure}
In Fig. 4 a-c, raw data are reported as a function of {\bf q} in $\AA^{-1}$, for {\bf q}//[110] (a), {\bf q}//[001] (b) and {\bf q}//[112] (c) at several temperatures. Here {\bf q} is defined with respect to 
the (110) Bragg peak.  Along [110], a modulation of I(q) appears very well-defined at 16K. It agrees with the observations reported close to \bmf{\tau}=(000) (Fig. 2), considering that slight differences may be expected due to the subtraction procedure. Along [112] and [001] a modulation of I(q) is also clearly seen at 16K, growing below the transition temperature, but with a smaller amplitude. Note the change of scale in both coordinates between Fig. 4-a and Fig. 4-b.
Along [001], beyond q=0.55$\AA^{-1}$, the increasing intensity corresponds to the tail of the
 short-range antiferromagnetic order as q approaches the AF (111) Bragg-peak.

By neglecting the displacement of Mn atoms, we can use the same 
analysis as that performed in section A. Since $S^{z\perp}$ in equation (1) is the spin component perpendicular to {\bf Q}= {\bf q}+\bmf{\tau} with \bmf{\tau}=(110), all directions are observable. Moreover, the corrections due to the geometrical factor can be neglected (cf Fig. 4-b).

All features indicate that the ferromagnetic correlations which define the ferromagnetic cluster 
have a smaller extension along [001] (antiferromagnetic stacking of the layers) 
 than along [110] (within the ferromagnetic layer). Assuming that all {\bf q} directions within the ferromagnetic layers are equivalent, such an anisotropy defines a plate-like shape for the inhomogeneities or that of an ellipsoid of revolution with main axes $\xi_{//}$, $\xi_{//}$, $\xi_{\perp}$. Since few directions have been studied, the q-dependences have been fitted
 independently with a gaussian function as an approximation of the F(q,R) function used in section A. Along [001], the q dependence beyond q$_{max}$ determines
 a typical size $\xi_{\perp}$$\approx$7$\AA$ ($\approx 2$ Mn-Mn distances), nearly twice as small as 
the value $\xi_{//}$$\approx$17$\AA$ found along [110]. The origin of this anisotropy will be discussed below, in relation with the anisotropic double exchange due to the AF layered structure.

In addition, the q$_{max}$ value at the maximum of intensity is larger along [001] (q$\approx$ 0.25$\AA^{-1}$) than along [110] ($\approx$ 0.18$\AA^{-1}$). We deduce an approximate inter-cluster distance $d_m$=2$\pi$/q$_{max}$
 smaller
 along [001] ($\approx$ 25$\AA$), than along [110] ($\approx$38$\AA$).

The behaviour with temperature has been mainly studied for {\bf q} along [110] (see Fig. 4-a). As T increases, a decrease of the intensity is observed.
 
The decrease of the elastic scattering does not result from a dynamical broadening. This was checked by energy scans at one constant q value, using the energy resolution 40$\mu$eV (using an incident neutron wavevector k$_i$=1.1 $\AA^{-1}$). Therefore, the signal remains "elastic" , even for our best energy resolution up to T$_{K}$. Close to T$_{K}$ a large scattering intensity enters in our finite experimental window, as shown in Fig 4-a at T=130K. It can be fitted by a lorentzian function, and is attributed to critical scattering. Above T$_{K}$ a residual bump of intensity persists, centered at q=0.  
In conclusion, the scattering intensity attributed to ferromagnetic polarons vanishes by approaching
 T$_{K}$ as does the ferromagnetic Bragg-peak intensity. The evolution of the q-dependence with temperature is difficult to determine, except at small q
 where a shift of q$_{max}$ to q=0 is observed. 
The residual scattering above T$_{K}$ suggests that some electronic segregation, which, in the temperature range T$<$T$_{K}$ gives rise to the ferromagnetic clusters, could also persist above T$_{K}$. Actually, such a persistence can be better studied from the temperature evolution of the unusual spin dynamics, as was reported for the Ca substituted samples\cite{Moussa2}.  
 
\begin{figure}
\centerline{\epsfig{file=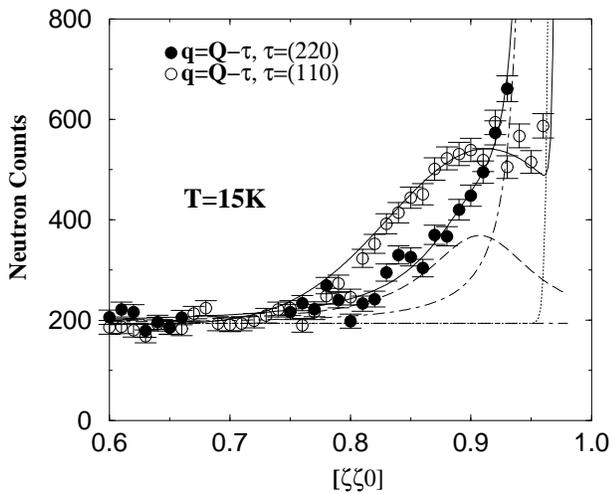,width=8cm}}
\caption{Comparison between diffuse scattering observed along [110], around $\tau$=(110)
 (empty circles) and (220) (full circles) using $k_i$=2.662$\AA^{-1}$. Close to $\tau$=(110), only a modulation fitted by a Gaussian (continuous line) is observed with the Bragg peak (dotted line). Close to $\tau$=(220), a new component (dotted dashed line) is observed in addition to the
 magnetic modulation (dashed line). }
\end{figure}

Finally, in Fig 5, the diffuse scattering along [110] observed around \bmf{\tau}=(110) and \bmf{\tau}=(220)
 are compared, both being measured in the same experimental conditions ($k_i$=2.662$\AA^{-1}$). The modulation does not readily appear around $\tau$=(220). Actually, besides the contribution of the Bragg peak, I(q) consists of a q-modulated intensity, twice smaller than that observed around $\tau$=(110) (dashed line), superimposed to a huge intensity at smaller q, approximated by a lorentzian function (dotted -dashed line). The decrease of the modulated intensity
 with {\bf Q} as \bmf{\tau} increases from (110) to (220), agrees with the decrease of the squared Mn $^{3+}$ magnetic form factor, ${\cal F}(Q)^2$ in equation (1), emphasizing the magnetic origin of the modulation. The lorentzian contribution, since not observed around $\tau$=(110), has no magnetic origin. It reveals a deviation from the perfect atomic long range order. Deviations from atomic positions give rise to a nuclear intensity varying with $Q^2$, whatever their isolated or segregated character\cite{Dederichs}. We recall that diffraction measurements performed on a very large Q range, and using a pair distribution function have determined two distinct Mn-O distances attributed to Mn$^{4+}$
and Mn$^{3+}$\cite{Louca}, and therefore a disorder for the oxygen positions. The present observation
 could be related to this disorder. 

\subsection{Antiferromagnetic short-range order}
As a feed-back effect of the ferromagnetic inhomogeneities described above, we also expect to observe some AF short-range correlations within the canted AF layered structure.
 Elastic diffuse scattering has been measured along [001], close to the antiferromagnetic Bragg-peak {\bmf\tau}=(001) at T=16K and T$>$T$_N$ (T=200K) and reported in Fig. 6. There, due to the geometrical factor, only correlations of spin components transverse to the [001] direction (S$^x$, S$^y$) are measured. Actually, below T$_{K}$, a diffuse scattering intensity grows. At 16K, the intensity consists of a weak modulation in the same q range as in Fig. 4-b, superimposed to a q-lorentzian intensity at the foot of the Bragg-peak.  This latter reveals an additional magnetic disorder, corresponding to a weak deviation from the AF ordering on a scale larger than the ferromagnetic cluster ($\approx$130$\AA$ from the q-linewidth). It likely corresponds to some rotation of the transverse components in the ({\bf a, b}) plane around their mean {\bf b} direction, which is an additional degree of freedom for these spin components. 
 \begin{figure}
\centerline{\epsfig{file=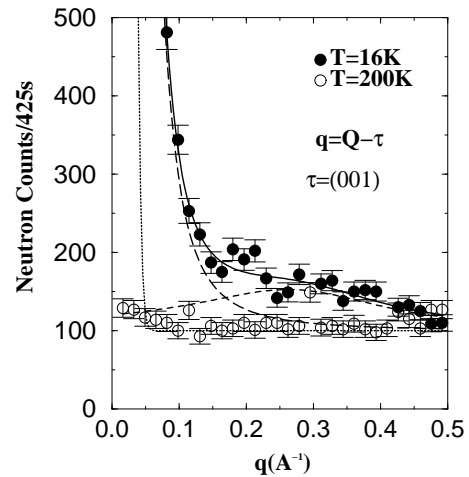,width=6cm}}
\caption{Scattering intensity versus q in $\AA^{-1}$ along [001], with {\bf q}={\bf Q}-$\tau$, $\tau$=(001): filled circle: T=16K empty circle: T=200K. Data at 16K are 
fitted by the sum of a delta function (dotted curve) for the Bragg peak with a lorentzian (long-dashed line) and a gaussian (small-dashed line) functions. 
 }
\end{figure}

\section{SPIN WAVES}

In the previous section, we have described an inhomogeneous state from a phenomenological
 point of view: cluster size, shape anisotropy etc. The determination of the magnetic excitations 
allows a better understanding of this peculiar magnetic state. Moreover, in the absence of twinned domains, the magnon intensities can be precisely determined, bringing new important informations.
Since we have already studied them in the Ca-doped samples, we can get a quantitative comparison between 
the two systems.

\begin{figure}
\centerline{\epsfig{file=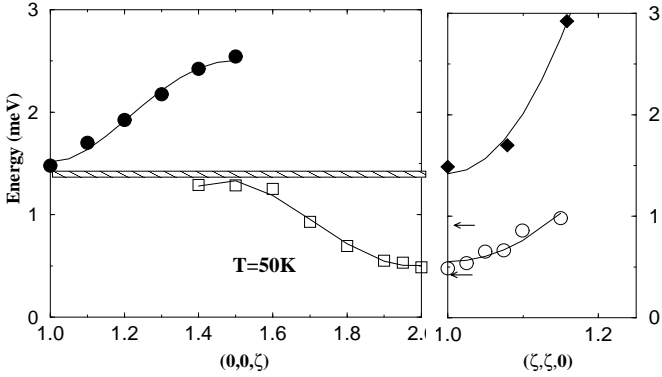,width=8.8cm}}
\caption{Dispersion curves of spin waves for q along [001] and along [110]. Filled symbols correspond to the high energy spin wave branch, and empty ones to the low energy one. The 
continuous lines are guides for the eye. The arrows point the gap value of the low-energy branch for
 x$_{Ca}$=0.05 (upper one) and  x$_{Ca}$=0.08 (lower one).}
\end{figure}

Inelastic neutron scattering experiments were performed
 on the spectrometer 4F1, at constant final neutron wavevectors k$_f$=1.55$\AA^{-1}$ and k$_f$=1.3$\AA^{-1}$ at 50K and 15K. 
Energy spectra for excitations in creation mode process have been fitted
by a lorentzian function as spectral function, convoluted with the spectrometer resolution:

$$S({\bf Q},\hbar \omega)=A{\cal F}(Q)^2 \hbar \omega (1+n(\hbar \omega))\chi({\bf Q})\times$$
\begin{eqnarray}
\frac{\Gamma({\bf q})}{(\hbar \omega-\hbar \omega({\bf q}))^2 +\Gamma({\bf q})^2}\frac{1}{\pi} \end{eqnarray}

 From the fits, the energy modes $\omega$({\bf q}), their damping $\Gamma$({\bf q}) and the energy integrated intensity $\chi({\bf Q})$ are determined. This latter parameter contains the geometrical form factor and any {\bf Q} dependence related to the symmetry of the magnetic structure. Since the modes are underdamped, the choice of the spectral function does not affect the results.
 $n(\hbar\omega)$ is the Bose population factor.

To benefit from the increase of intensity related to the thermal population factor, the two
 main-symmetry directions [001] and [110] have been fully studied at 50K, whereas only the [001] direction has been studied at 15K. 
 
In Fig. 7, $\hbar$$\omega (q)$ are reported for $q$ in $rlu$ along [001] and [110] directions at T=50K, focusing on the energy range below 3 meV. The scale in abscissa has been chosen to allow a direct comparison 
of the two directions in $\AA^{-1}$ units. 
The similarities of the dispersion curves with previous observations in Ca-substituted samples are striking\cite{Hennion1,Moussa2}: two dispersive spin-wave branches are observed. One with a large gap and an anisotropic dispersion, characteristic of the AF layered structure. Corresponding spectra are shown in Fig 8-a for direction [001].The other one, with a small gap and an isotropic dispersion, is observed at nuclear Bragg-peaks only. It is characteristic of a ferromagnetic coupling. Along [001], it levels up at the zone boundary (001.5), and lies below the gap of the former (see the hatched area along the [001] direction in Fig. 7). Corresponding spectra are shown in Fig. 8-b, and 8-c for directions [001] and [110] respectively.

By fitting the high energy branch along [001] with a dispersion relation deduced from a Heisenberg 
hamiltonian as in 
the pure system\cite{Moussa1}, we determine an interlayer "effective"
 antiferromagnetic coupling ($J_2$=-0.3 meV at T=15K), strongly reduced with respect to the undoped system\cite{Moussa1} ($J_2$=-0.58 meV at T=15K).  Along [110], high energy measurements are lacking to determine the $J_1$ coupling, which was found to increase slightly with x in Ca-doped samples\cite{Moussa2}.

 The low-energy spin wave branch has been analyzed with a phenomenological quadratic law at small q, $\hbar$$\omega$=$Dq^2$ +$\hbar$$\omega_0$
 ($D=6JSa_0^2$, if J is the first neighbor coupling). The fit determines 
$\hbar$ $\omega_0$=0.54$\pm$0.02 meV and $D$=10 meV$\AA^2$ at 15K in both directions. 

$\Gamma$({\bf q}), which characterizes the excitation damping of the low-energy branch, obeys well
 a quadratic law, 
identical for the [001] and [110] directions (see Fig. 9 for the [001] direction).  

The {\bf Q} dependence of the energy-integrated intensity, $\chi ({\bf Q})$, reveals several new features.

1) The  q-dependence  of the low energy branch, measured in the same Brillouin zone
 (centered at \bmf{\tau }=(002)) strongly differs for {\bf q}//[001] and for {\bf q}//[110].
 This is illustrated in Fig. 10, where the two {\bf q} directions
 are compared. 
 The fit with a q lorentzian lineshape (continuous lines in Fig. 10) yields correlation ranges along [110] and along [001] of 
19$\AA$ and 4 $\AA$, respectively. This
q-dependence is reminiscent of the ferromagnetic correlations $\xi_{//}$ and $\xi_{\perp}$ obtained for the static ferromagnetic clusters. Because of the 
strong decrease of the intensity along [110], the low-energy spin wave branch could not be measured beyond some 
q value ($\zeta$= 1.15 in Fig. 6), which is far from the zone boundary ($\zeta$= 1.5) along this direction.

(2) The intensity of the magnetic excitations of the low energy branch is twice larger for measurements in the Brillouin zone centered at \bmf{\tau }=(002), than in that centered at \bmf{\tau }=(110). This is illustrated in Fig. 8-d, for one q value. 
Interestingly, this factor $2$ corresponds to the geometrical factor expected for spin fluctuations transverse to the {\bf c} axis, namely, the direction of ferromagnetism in this sample. This finding agrees with the assignment of the low-energy spin-wave branch to the new ferromagnetism induced by doping.
We mention that, in our previous study of Ca-substituted samples\cite{Moussa2}, the reported measurements have suggested a different conclusion (intensity larger around (110) than (002)). This difference is likely not real, but related to the difficulty of separating the domain contributions in twinned samples.

(3) Finally, we outline the absence of any intensity for magnetic excitations of the high energy branch
 at \bmf{\tau}=(002). This is illustrated in Fig. 10, where the full squares show the decrease of the spin wave intensities of the high energy branch from (001) to (002) along the [001] direction. This variation
is a consequence of the symmetry of the antiferromagnetic layered structure. 
 Therefore, we can assign the high energy branch to spins (or spin components) with the antiferromagnetic A type structure. Accordingly, along [001], the intensity of the high energy branch, maximum at (001) (full squares) and the intensity of the low energy one, maximum at (002) (open circles) have complementary q-dependences. At intermediate q, both types of excitations can be observed (see $\zeta$=1.4 and 1.5 in reduced 
units in Fig 8-a).

The spin dynamics of the x$_{Sr}$=0.06 sample has also been compared to that of the x$_{Ca}$=0.05 and 0.08 samples.  

 In Fig. 11, the dispersion of the low energy branch along [001] is reported for x$_{Ca}$=0.05 (open squares), x$_{Ca}$=0.08 (open circles) and x$_{Sr}$=0.06 (full circles), 
all measurements being performed  at T=15K.

\begin{figure}
\centerline{\epsfig{file=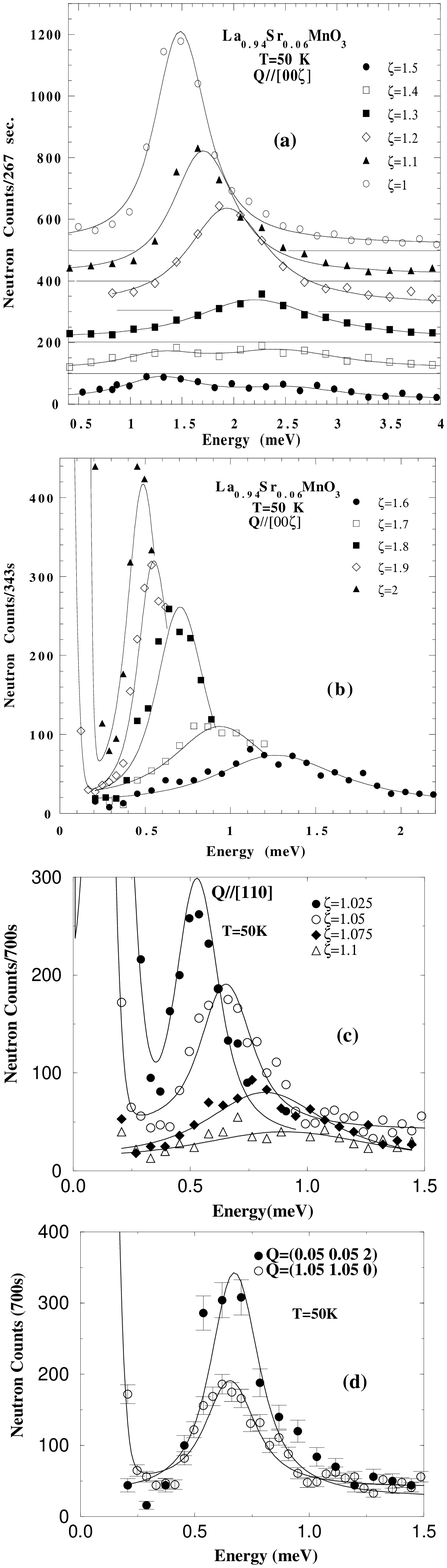,width=6.5cm}}
\caption{Energy scans at constant {\bf Q} {\bf a)}: along [001]
 at  k$_f$=1.55$\AA^{-1}$, {\bf b)}: along [001] at k$_f$=1.3$\AA^{-1}$,
 {\bf c)} along [110] with $k_f=1.3\AA^{-1}$,  {\bf d)} along [110] 
in two different Brillouin
 zones. The continuous
 lines are fits with a lorentzian function convoluted with the spectrometer resolution. 
 }
\end{figure}

 \begin{figure}
\centerline{\epsfig{file=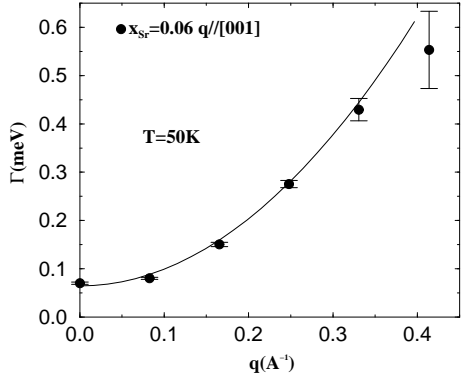,width=6cm}}
\caption{Linewidth of the low-energy spin wave modes $\Gamma$ versus q for q along [001] at T=50K.}
\end{figure}
\begin{figure}
\centerline{\epsfig{file=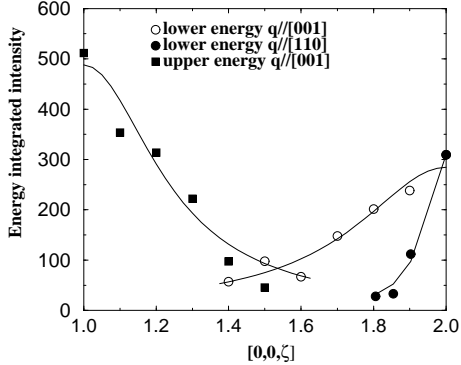,width=6cm}}
\caption{q dependence of the magnon peak intensity at 50K: along [001] for the high energy branch (filled squares)
 and the low energy branch (empty circles) and along [110] for the low energy branch starting from (002), and reported in equivalent units of $rlu$ [001]  (filled circles).The continuous lines are a guide for the eyes.}
\end{figure}

\begin{figure}
\centerline{\epsfig{file=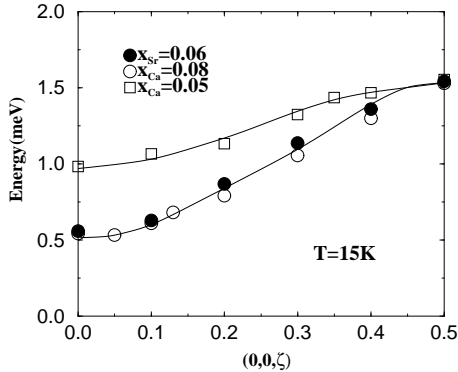,width=6cm}}
\caption{Dispersion of the low energy branch along [001] at T=15K for
 x$_{Ca}$=0.05 (empty squares) x$_{Ca}$=0.08 (empty circles) and x$_{Sr}$=0.06 (full circles).}
\end{figure}

This comparison reveals that
the low energy branch of x$_{Sr}$=0.06 corresponds very precisely
to that measured in the x$_{Ca}$=0.08 sample,
 well distinct from that in the x$_{Ca}$=0.05 sample. For instance, the gap value $\omega_0$=0.55 meV, is twice smaller
 than the value (0.91 meV) found at x$_{CA}$=0.05.
  The same shift with x can be seen, considering the value determined by EPR measurements\cite{Ivanov}, namely $\omega_0$=0.74 meV for x$_{Sr}$=0.05, significantly smaller than that measured at the same hole concentration in the Ca-substituted sample (0.91meV). 
This enhancement of the ferromagnetism upon doping with Sr can also be seen when considering
  the antiferromagnetic coupling ("effective" magnetic coupling $J_2$) deduced 
from the high energy branch, which is the same as that found for x$_{Ca}$=0.08.  

 We conclude that the characteristics of the low energy branch (gap and ferromagnetic coupling), as well as those of the high energy branch, evolve faster with x for Sr doping than for Ca doping. Moreover, the shift with x ($\approx 2$\%) is the same when comparing the spin dynamics and the evolution of the static ferromagnetic clusters (section III-B). Therefore, a relation exists between the size of the ferromagnetic cluster and the ferromagnetic coupling or the gap value defined by the low-energy spin wave branch.

In this comparison, we also note that the large gap value of the high energy branch, 1.86 meV at T=15K, is, within our accuracy, the same as the gap value that was determined in Ca-doped samples, which is constant in the $0.05\le x_{Ca} \le 0.1$ concentration range\cite{Hennion2,Moussa2}. Correspondingly, the value of the low energy branch at the zone boundary appears also fixed (see at (001.5) in Fig. 11), so that the spin wave lies below the
 gap of the high energy branch, in a separated energy range.

Preliminary results in LaMnO$_{3+\delta}$ confirm the general character of these observations.

   \section{Discussion and conclusion}

In this paper, we have shown that all the unusual static and dynamical features found in the Ca 
doped samples,
 also exist with Sr doping, confirming the general character of our previous observations. 

The analysis which takes into account the anisotropy of the spatial distribution, indicates a density of ferromagnetic clusters $\approx$ 25 times smaller than the hole density. This suggests a picture of an electronic phase segregation with hole-rich regions, as a general feature of the low doping state. The absence of any chemical segregation, checked by X-rays for one Ca-doped sample
 (x=0.08), confirms this electronic origin. 
However, the  ferromagnetic component deduced from the Bragg-peak intensities cannot be 
explained 
by the volume fraction of ferromagnetic clusters only, suggesting that a small ferromagnetic component also exists within the matrix.

The two new results are: 1) anisotropy of the ferromagnetic cluster shape, indicating platelets rather than droplets, 2) anisotropy of the dynamic susceptibility, which contrasts with the isotropy
 of the dispersion curve. Both features have been observed thanks to the absence of twinning in
our sample. 
 
The anisotropy of the ferromagnetic cluster shape must be a consequence of the AF layered magnetic structure and of the double-exchange coupling of Zener\cite{Zener} which favors the hole hopping 
within the ferromagnetic layer. 
The layered structure is related to the cooperative Jahn-Teller effect which lifts the degeneracy of the $e_g$ state\cite{Feiner,Feinberg}. 
The optimal size of the cluster can be defined from the competition between the kinetic 
energy of the e$_g$ hole 
which is favored
 by the growth of the ferromagnetic cluster, and the long-range coulomb energy, which unfavors the 
electronic segregation\cite{Nagaev}.  The enhanced ferromagnetic character induced by the Sr dopant with respect to the Ca dopant, confirms the r\^ole of some structural parameter for determining the hopping probability or the kinetic energy.

Turning now to the spin dynamics, we observe a coexistence of two spin wave branches, the high energy one, typical of superexchange coupling, and the low energy one indicating a
 ferromagnetic and isotropic coupling. From the spin-wave intensities observed at the (002) and (110) Bragg peaks, we can assign the high energy branch to spins (or spin components) with the AF A-type structure, and the low energy one to spins (or spin components) ferromagnetically aligned along {\bf c}. 

The fast decrease of the AF inter-layer coupling along [001], defined by the high-energy spin wave branch, and assigned to superexchange coupling, agrees well with previous experimental results\cite{Hirota}. This evolution characterizes an increasing ferromagnetic character, showing that the spin coupling , typical of the antiferromagnetic layered structure in which the clusters are embedded, is very sensitive to the doping x. Specially, this variation with x, can 
be explained within a model\cite{Feiner}, which does not assume any electronic phase segregation and does not consider the canted state.
  
However, such a model cannot explain the low-energy spin-wave branch. The interesting result, concerning this low-energy spin wave branch, is the perfect isotropy of the corresponding dispersion (and of the  damping), within this AF layered magnetic structure. Moreover, the q-dependent susceptibility, strongly anisotropic,
 reflects closely the anisotropic shape of the ferromagnetic static clusters.
All these characteristics  lead to connect these magnetic excitations with the ferromagnetic hole-rich regions. 

The absence of any cut-off size effect or any anomaly at q$_{max}$ in the dispersion curve, in contrast with the case of the stripe picture for instance\cite{Tranquada}, prevents a true 
inhomogeneous picture. Rather, it suggest that these spin waves are coherent through the antiferromagnetic matrix. The long-range ferromagnetic order, found in the system (canted state) is certainly responsible for that. 
 Moreover, the origin of the small energy gap and its evolution with x, could be understood by using general concepts of second order phase transitions. This allows to make some predictions for the evolution with x of $\chi(q,\omega)$
in $\omega$-q space. We recall that, as x increases, the small gap energy at q=0 decreases and the strength of the ferromagnetic coupling defined by the q-dependence increases. 
 Therefore, the gap value at x could
characterize a "distance" towards a critical x$_c$ value at T=0. This critical concentration would correspond to 
a {\it uniform} ferromagnetic state. Within such an approach, the new dynamical susceptibility $\chi$ ($\omega$,q)
 can be described as a precursor effect of this "uniform" ferromagnetic and metallic phase. The isotropy of the new ferromagnetic coupling indicates that, in this region, the magnetic unit cell is the small perovskite cube, as in the metallic phase occurring at x$>$x$_c$. 

The new ferromagnetic coupling is likely induced by itinerant e$_g$ holes. Therefore, in spite of the macroscopic insulating character of the low-doped samples, band models\cite{Nagaev,Gor'kov,Kagan,Yunoki,Moreo,Yi}\cite{Millis}\cite{Ishihara} which consider itinerant $e_g$ holes in addition to localized spins, should be used. 

This precursor dynamics is expected to disappear at the metallic transition where the hole-rich and hole-poor regions melt into a single phase. 
Actually, at x$_{Sr}$=0.3, a single spin dynamics is observed with a vanishingly small gap value\cite{Martin}.
This will be discussed in a further paper on Ca substituted samples where the spin dynamics at the transition has been studied in a detailed way\cite{Biotteau2}.

Finally, we have outlined that, within our experimental accuracy, the large gap of the high energy branch and the magnon energy of the low-energy branch at the zone boundary (001.5) appear to be 
fixed at some general values. Moreover, the low energy branch lies below the gap of the high energy one (see the hatched area in Fig. 7), in a separated energy range.
 This latter peculiar feature has similarities with observations in La$_2$NiO$_{4.11}$\cite{Nakajima}, where it is discussed in terms of host-host and impurity-host couplings,
  by analogy with the single impurity problem\cite{Tonegawa}. It is worth-noting that, unlike the case of
 La$_2$NiO$_{4.11}$, the two spin wave branches appear in two distinct symmetry points in the reciprocal space (001) and (002), defining two magnetic coupling of opposite
 sign along {\bf c}. The competition between these two 
magnetic couplings could be responsible for this separation in energy. 

In conclusion, the static and dynamic spin correlations in low-doped manganites characterize a picture of two-magnetic regions, confirming an electronic phase segregation. However, the existence of a long range ferromagnetic order indicates that the ferromagnetic regions are connected through the matrix leading to a non uniform canted state. Moreover, the spin dynamics assigned to the AF spins or spin components, is strongly sensitive to the doping, acquiring a stronger ferromagnetic character as x increases. Therefore, some homogeneous and inhomogeneous characteristics are actually observed together, whereas they are exclusive in all the theoretical models. This leads to a weakly inhomogeneous picture. Its  true origin 
remains to be clarified.

\section*{Acknowledgements}

We are very indebted to P. A. Lindgard, F. Onufrieva, Y. Sidis, L. P. Gor'kov, S. Ishihara , A. M. O\'les
and M. Kagan for fruitful discussions.

\end{document}